\documentclass[submission,copyright,creativecommons]{eptcs}
\providecommand{\event}{GraMSec 2014} % Name of the event you are submitting to
\usepackage{url}
\usepackage[dvips]{graphicx}
\usepackage{caption}
\usepackage[font=footnotesize, caption=false]{subfig}
\usepackage{todonotes}

\usepackage{booktabs}

\title{Threats Management\\Throughout the Software Service Life-Cycle}

\author{Erlend Andreas Gj{\ae}re
\institute{SINTEF ICT\\ Trondheim, Norway}
\email{erlendandreas.gjare@sintef.no}
\and
Per H{\aa}kon Meland
\institute{SINTEF ICT\\
Trondheim, Norway}
\email{\quad per.h.meland@sintef.no}
}
\def\titlerunning{Threats Management Throughout the Software Service Life-Cycle}
\def\authorrunning{Erlend Andreas Gj{\ae}re \& Per H{\aa}kon Meland}

\begin{document}
\maketitle

%%% ABSTRACT %%%
\begin{abstract}
Software services are inevitably exposed to a fluctuating threat picture. Unfortunately, not all threats can be handled only with preventive measures during design and development, but also require adaptive mitigations at runtime. In this paper we describe an approach where we model composite services and threats together, which allows us to create preventive measures at design-time. At runtime, our specification also allows the service runtime environment (SRE) to receive alerts about active threats that we have not handled, and react to these automatically through adaptation of the composite service. A goal-oriented security requirements modelling tool is used to model business-level threats and analyse how they may impact goals. A process flow modelling tool, utilising Business Process Model and Notation (BPMN) and standard error boundary events, allows us to define how threats should be responded to during service execution on a technical level. Throughout the software life-cycle, we maintain threats in a centralised threat repository. Re-use of these threats extends further into monitoring alerts being distributed through a cloud-based messaging service. To demonstrate our approach in practice, we have developed a proof-of-concept service for the Air Traffic Management (ATM) domain. In addition to the design-time activities, we show how this composite service duly adapts itself when a service component is exposed to a threat at runtime.
\end{abstract}

\section{Introduction}
With the emerging paradigm of composite services, i.e. software services which are composed of functionality provided by several services components, possibly involving several service providers, an attack on a service component implies an attack on the composite service as a whole. Service compositions in general are highly distributed and have a complex nature, exposing a greater attack surface than traditional stand-alone systems. To make this kind of services secure enough, it is vital to take a holistic approach to how they are built, security-wise. This assumes  incorporation of activities that deal with threats at many stages throughout the life-cycle, comprising both design and development as well as the runtime phase.

At design-time, we can utilise preventive activities for mitigating threats, and at runtime there should be performed corrective activities to handle the residual threats. It is very optimistic to think that all threats are known and correctly assessed from the very beginning, e.g. it is not given that we have reliable knowledge of motivation and available resources for a potential attacker or enough resources to implement preventive measures. What we do know, is that when a threat escalates, we need to make sure that our organisation and the system(s) we are trying to protect are prepared to respond effectively.

Our main goal with this paper is to describe a process and tool-chain which combines threat modelling with graphical goal and process models to derive preventive measures, such as security requirements for development, and also enables further corrective measures to be automatically applied at runtime. We have chosen to support security analysis through activities and tools that are already adopted and in use, although these have not necessarily been previously combined. 

The details of our proposed approach is given in section~\ref{approach}, incorporating our Air Traffic Management system case study. Through this we explain and demonstrate the use of a threat repository to facilitate sharing and re-use of threats, goal-oriented modelling with threats, service design with BPMN, model transformation, threat response recommendation, defining rules for dealing with escalations, and finally runtime management of composite services. Section~\ref{discussion} discusses strengths and compromises made in this approach, and section \ref{conclusion} concludes the paper.

%\todophm{figure showing threats in the lifecycle, risk analysis during requirements, design flaws, technology dependent threats during implementation, test identification, maintenance and updates based on new threats}

% Edit separate file introduction.tex

\begin{figure}[htb!]
    \centering
    \includegraphics[width=1\textwidth]{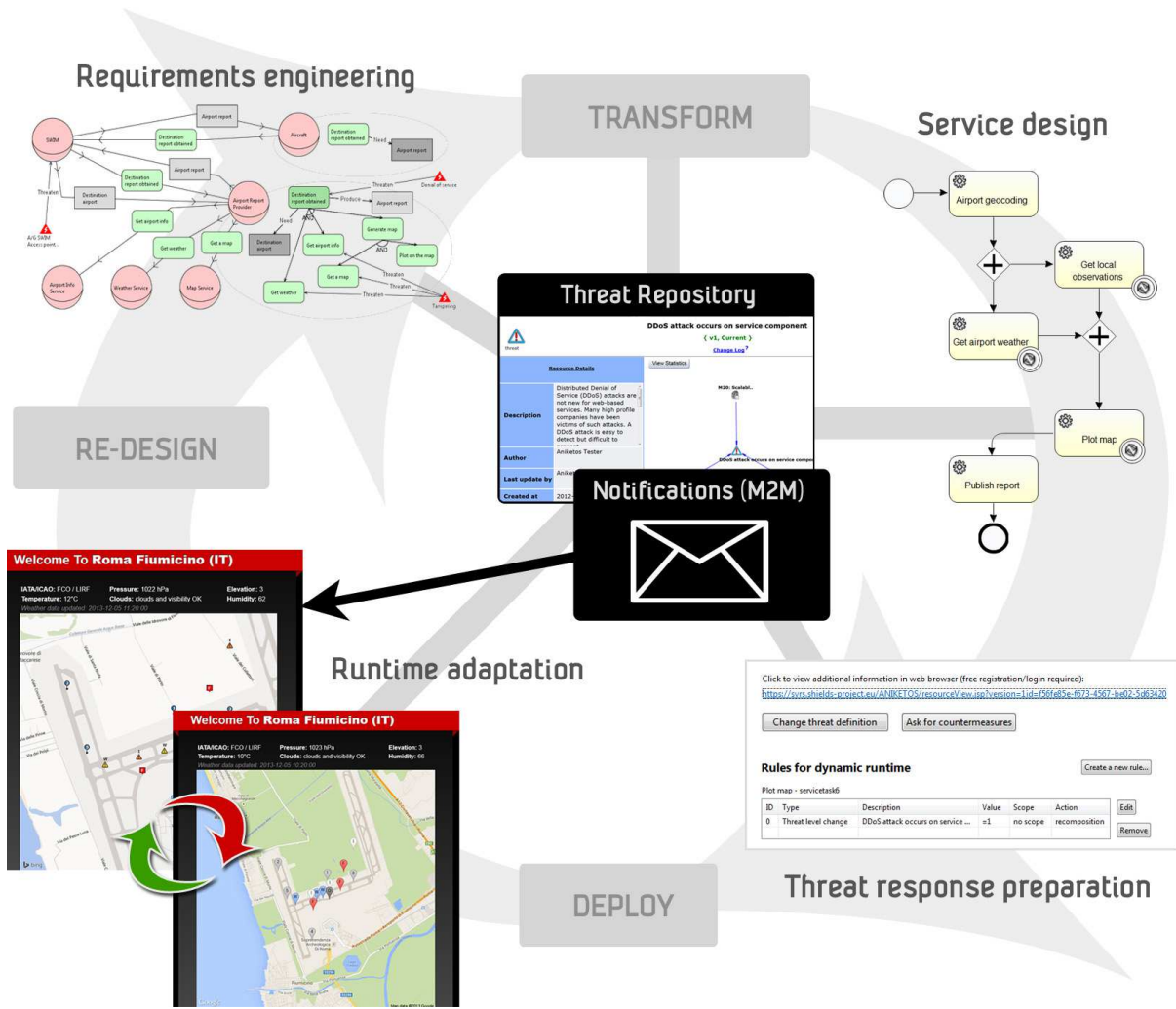} %,bb=0 0 1635 1393
  \caption{A software life-cycle model showing how threats can be modelled and managed at design-time and prepared for automated service adaptation at runtime.}
\label{fig:lifecycle}
\end{figure}

\section{Approach Description}
\label{approach}
Figure~\ref{fig:lifecycle} gives a visual overview of the threat management life-cycle, outlining both activities in the process as well as data flow. Key to understanding the advantages of this holistic view are the threat repository and notification services in the centre, which together connect the dots between the remaining components and stages. In the requirements engineering process we utilise goal-oriented modelling (Socio-Technical Security modelling language, STS-ml~\cite{ststool}) to elicit and define security requirements. Here we also instantiate some of the main threats to our system on an organisational/business level of abstraction. Further, we use Business Process Model and Notation (BPMN)~\cite{OMG} for designing the service process of a composite web service. At this stage we do not depend upon how the requirements engineering has been done, but if STS-ml has been used, we can transform the goal models into process models with some help from a software tool. The BPMN model defines a service process which can be deployed and executed as a web service, and here should the threats be refined on a more technically detailed level. Based on this, we can ask for accordingly technical countermeasures from the threat repository. We can also define rules for responding to particular events, if we are able to monitor the service components at runtime. Whenever an alert is received from the monitors, these rules will be evaluated by a service runtime environment (SRE) tailored for this purpose. If is a match is found, a re-composition can for example be triggered, replacing a service component instance with another one providing similar functionality. Alternatively, one can move along in the life-cycle to re-design of the system and/or the particular service.

\subsection{Case Study: Air Traffic Management}
For demonstrating our proof-of-concept, we have chosen a case study based on Air Traffic Management (ATM). The European air navigation services and their supporting systems are currently undergoing a paradigm shift, most notably through the System Wide Information Management (SWIM)~\cite{swim}. Going from numerous incompatible stand-alone systems, SWIM enables cross-border collaboration and data exchange between many systems and organisations with its service oriented architecture (SOA) approach. This raises prospects for an expansive registry of composite services, which hopefully will contribute to maximizing efficiency of the airspace with time. The service we demonstrate is one that gathers various information about an airport to be used e.g.\ in the cases when a pilot wants a fresh report on the flight's destination. As seen from the perspective of this service provider, further details on this composite service sample is provided along with the description of each step in our approach below. Security-experts from three ATM organisations were involved in the modelling, using already defined security requirements and system specifications from the SESAR project~\cite{sesar} through which SWIM is developed. However, the models provided in this paper only represent a limited excerpt of the SWIM system model, and the demonstrator we have implemented is not based on service components actually provided through SWIM. Work on collecting and analysing feedback from the actual case study users is yet to be completed, and results as such are hence not considered in this paper.

%%% STRUCTURE FOR SUBSECTIONS
% - How this step is intended for the approach in general
% - Explanation of case study specifics
% - Technical details of our PoC for this step
% - (Additional notes on the threat(s))

\subsection{Knowledge Management Through a Threat Repository}
Every part of our approach is tied together by the concept of having a centralised repository of threats~\cite{5066571}. While threats can always be created and modelled independently in the different diagram types that we use, a persistent threat repository provides better conditions for import and re-use of threats~\cite{Oladimeji06securitythreat}. It is of essence here that each threat in the threat repository is associated with a unique threat identifier (ID), and that this threat ID is stored along with the threats in the various diagrams we create. This allows the tools to always access (meta-) information on threats stored in the repository, and for the threats to be passed between the models and pushed further through deployment into the runtime phase.

\begin{figure}[htb!]
    \centering
    \includegraphics[width=0.8\textwidth]{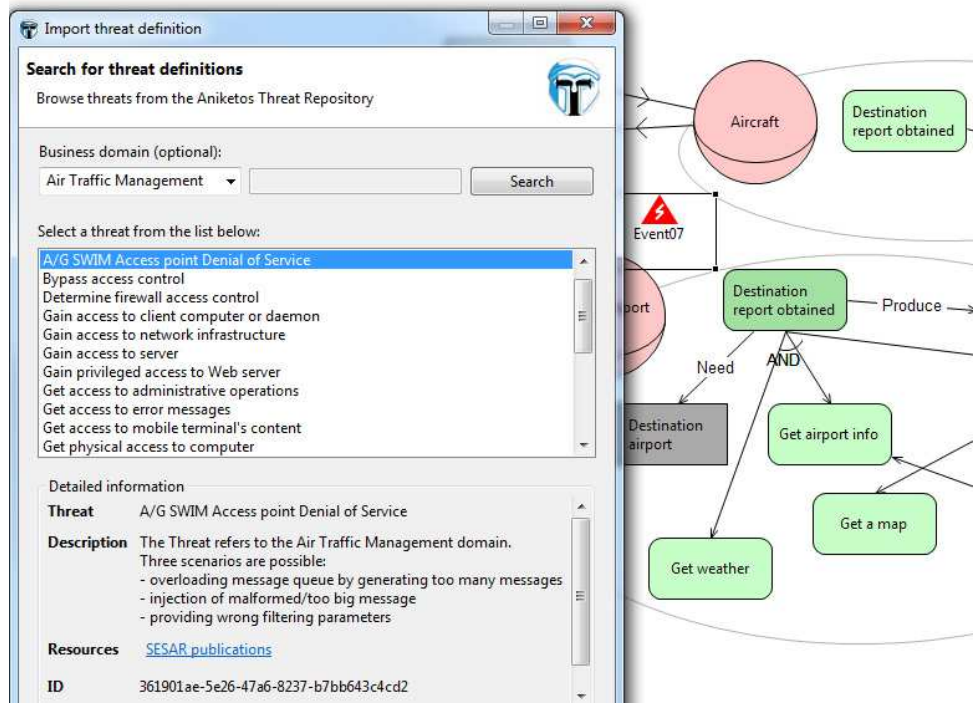} %,bb=0 0 775 570 
  \caption{Importing a domain-specific threat from the threat repository into a graphical model.}
\label{fig:import-threats}
\end{figure}

In our proof-of-concept, we have utilised an existing online threat repository service~\cite{svrs}, providing an application programming interface (API) for accessing its large collection threats externally. The API offers functionality for searching for threats in terms of their name, class (business-level or operational level), or business domain tags. Some additional meta-data may also be available, such as a textual description and links to further information resources. For the two graphical modelling tools we present below, we have implemented a plug-in that takes advantage of this API, as shown in Figure~\ref{fig:import-threats}. In addition to the API, we have access to a web-based user interface which allows more info to be looked up on each threat, including suggestions for countermeasures and relationships with other threats. A threat uploader tool has also been developed for adding new threats to the threat repository, using existing diagrams.

Threats related specifically to ATM are found in the threat repository under the business domain selector \textit{Air Traffic Management}. These threats can both be general ones that are known to also apply in that particular domain, such as \textit{Gain access to server}, as well as threats more specific to ATM only -- e.g.\ \textit{A/G SWIM Access Point Denial of Service}. Generic high-level threats, like \textit{Tampering}, can for instance be specialised into a domain specific threat such as \textit{False airport coordinates}.

\subsection{Requirements Engineering}
While requirements engineering can take many shapes and forms, graphical goal-oriented modelling is an approach well suited for complex and distributed socio-technical systems (of systems)~\cite{sts2, PajaDPRG2013}. Where goals justify why the system and its functions are needed, threats justify why \textit{security} for the system is needed.

\begin{figure}[htb!]
    \centering
    \includegraphics[width=1\textwidth]{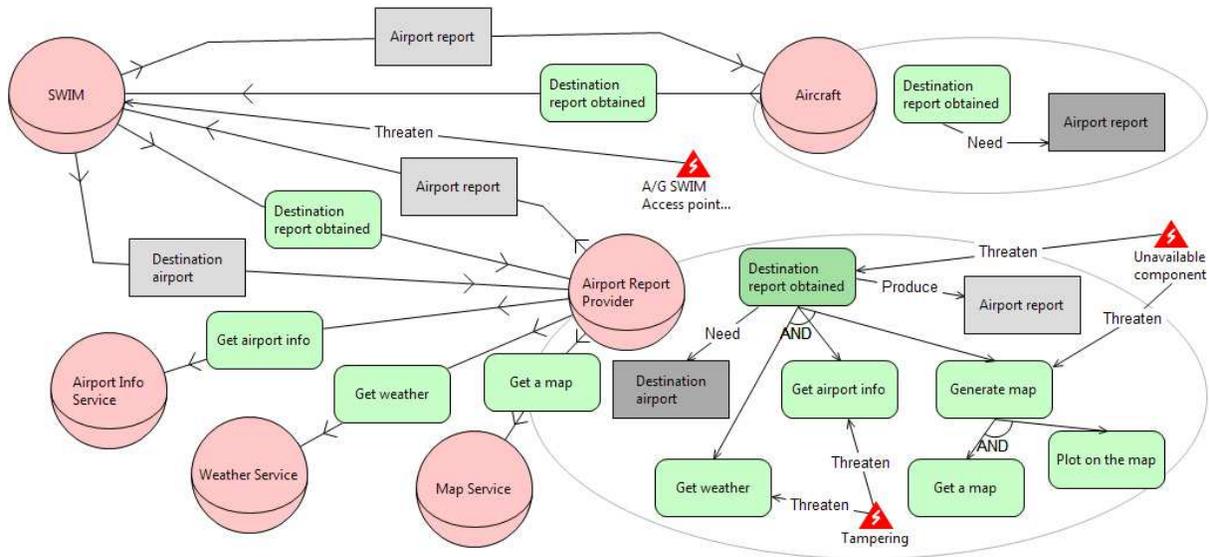} %,bb=0 0 1028 476
  \caption{Goal-oriented model (STS-ml) of the airport report service -- the red triangles are threats.}
\label{fig:sts-social}
\end{figure}

In the STS model from our case study, depicted in Figure~\ref{fig:sts-social}, we see that the pilot's request for a destination report is transmitted to SWIM rather than directly to a specific service provider. Since the requesting aircraft's destination is information already known inside SWIM, the request is complemented with this information here. Through its service registry, SWIM can select and invoke an actual service provider on behalf of the pilot/aircraft, possibly depending on the particular destination's location. The airport report contains various local information on the airport, along with weather conditions and a map onto which various local observations (e.g.\ wind meter readings, contaminations) are plotted. Threats from the perspective of the airport report provider, would in this case be any kind of tampering with externally acquired data or unavailability of any of the service components, which might hold them accountable for impacting flight safety. A threat worth including could also be denial of service within the SWIM area of responsibility, since that could block an airport report to arrive timely on the provider's behalf -- although not being the airport report service provider's actual responsibility in the end.

\begin{figure}[htb!]
    \centering
    \includegraphics[width=0.8\textwidth]{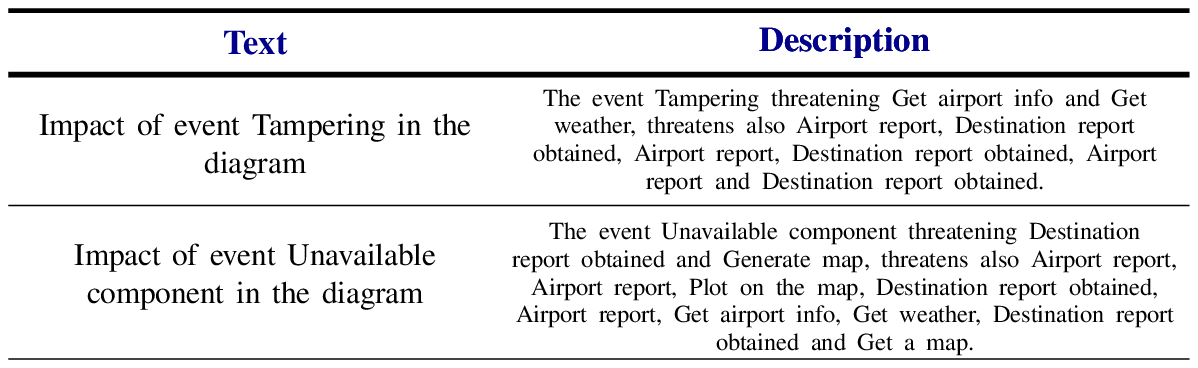} %,bb=0 0 1484 456
  \caption{Propagation of threats can be analysed with the generated security requirements document.}
\label{fig:doc-threats}
\end{figure}

Supported by the STS-Tool~\cite{ststool}, we are able to both perform both graphical modelling as well as advanced formal analyses of the models. This includes checking consistency and detecting conflicts between goals and requirements, as well as analysing (visually in the model) how threats can propagate throughout the modelled system~\cite{goals-threats}. The STS-Tool also supports generating a security requirements document which contains the information we have added on threats. In addition, the document generator performs the aforementioned threat propagation analysis so that the results can be output in textual form, as shown in Figure~\ref{fig:doc-threats}. In our case specifically, the \textit{A/G SWIM Access Point Denial of Service} threat does not have a propagated impact, and is hence not mentioned in the document, as it is directed at an actor external to the modelled system owner. Although not implemented, the document generator could potentially use the threat IDs provided in the model to look up information on countermeasures, and hence include that in the document for further reference.

\subsection{Service Design}
In the BPMN model, we specify the process flow which our composite service shall follow. Herein lies the advantage of standardised BPMN as the service process language, that the graphical models translate into well-defined execution semantics, which in turn can be executed in business process model engines as a web service. Although BPMN has no explicit language construct for threats, we have in previous work concluded that e.g.\ the standard \textit{ErrorBoundaryEvent} element can be used for representing threats~\cite{conf/IEEEares/MelandG12}. %dd threats further to enrich the process flow on a technically detailed level.  

\begin{figure}[htb!]
    \centering
    \includegraphics[width=1\textwidth]{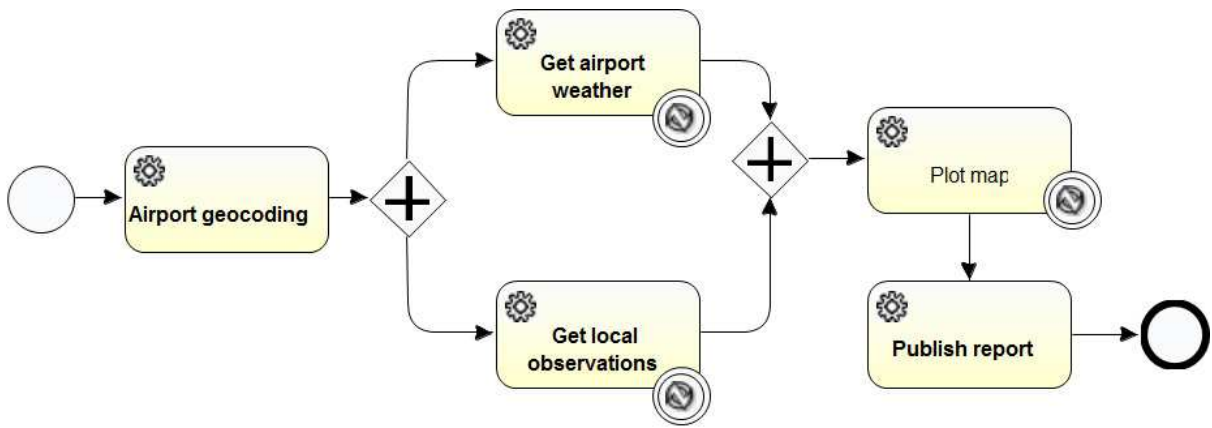} %,bb=0 0 924 321 
  \caption{Service process specified in BPMN, using standard ErrorBoundaryEvent elements for threats.}
\label{fig:bpmn}
\end{figure}

While our BPMN model in Figure~\ref{fig:bpmn} may appear similar to the STS model, the service tasks (yellow boxes) now represent atomic components of a process, unambiguously ordered as an executable process flow. Moreover, each service task implies a single invocation of a particular web service, taking process variables as (optional) input to operations defined by each service components' web service definition file (Web Service Definition Language, WSDL~\cite{w3cWSDL}). Although not visible in the model, the first service task, \textit{Airport geocoding}, takes as input the IATA code of an airport (e.g.\ \textit{FCO} for Rome, Fiumicino) and queries a public airport information service. The returned value, which is also defined by the component's WSDL file, should be a pair of coordinates which pinpoint the airport's geographical location. These coordinates are in turn used as input for the following two tasks, which can be done in parallel, namely to obtain the weather and any local observations from/surrounding the location in question. A fourth service takes care of plotting the gathered information on a map, whereas the final (internally maintained) service wraps up and creates the report data to be returned to the pilot's application.

\begin{figure}[htb]
    \centering
    \includegraphics[width=0.8\textwidth]{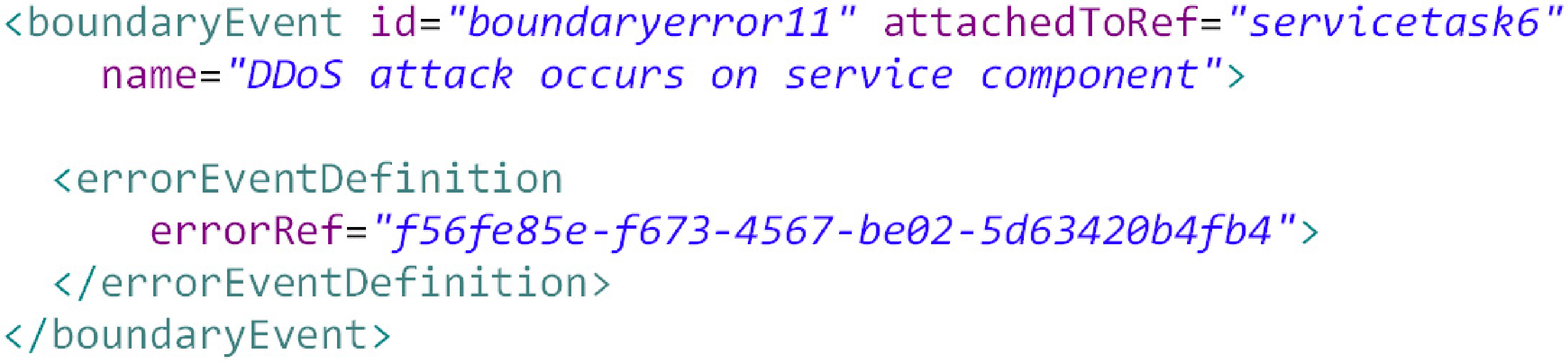} %,bb=0 0 1086 253 
  \caption{Using an ErrorEventDefinition XML-element with errorRef attribute to store the threat ID.}
\label{fig:errorevent-xml}
\end{figure}

We have extended the Activiti Designer tool~\cite{activiti}, built as a plug-in for the Eclipse IDE~\cite{eclipse}, in order to support our BPMN modelling. We have chosen to store the threat ID in the BPMN 2.0 XML file, as shown in Figure~\ref{fig:errorevent-xml}, which is in line with the standard. This results in compatibility between different modelling tools, at least when it comes to portability of the graphical model. What is not part of BPMN is functionality for producing several alternative composition plans, i.e. alternative combinations of service components, that each provide the very same functionality~\cite{6488561}. This is possible when you have more than one possible candidate services for a service task, while still maintaining the required level of functionality. In our proof-of-concept, we have two candidates for the map service. The total number of composition plans equals the Cartesian product of all service component candidates for all service tasks, i.e. all possible combinations of service implementations. These different composition plans can in turn be ranked against one or several criteria, such as level of trustworthiness, quality-of-service, etc., if this information is maintained on behalf of the components~\cite{achim, brucker.ea:framework:2013}.

\subsubsection{Model Transformation}
\begin{figure}[htb!]
    \includegraphics[width=0.8\textwidth]{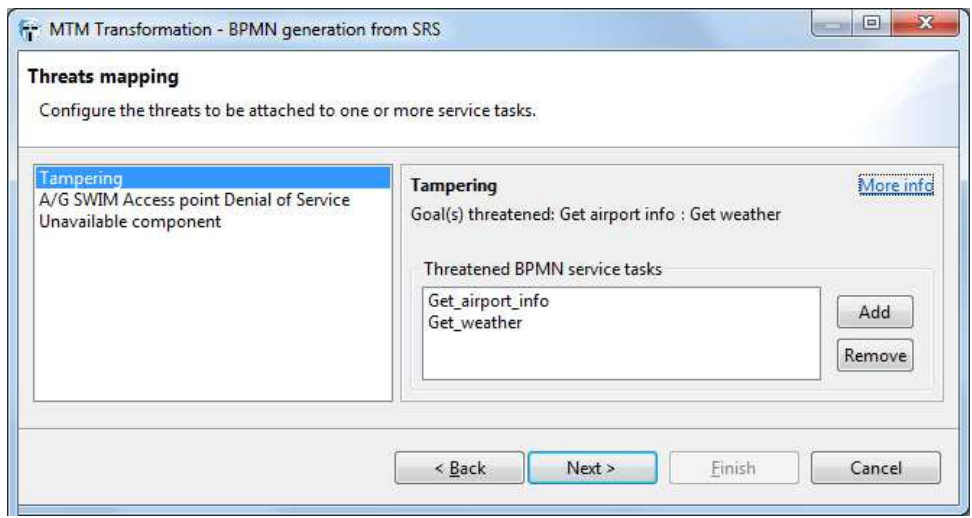} %,bb=0 0 666 352 
    \centering
  \caption{Tool-supported model transformation of threats from STS-ml (via SRS) to BPMN.}
\label{fig:mtm-transformation}
\end{figure}

In order to support security information from the requirements engineering phase to be maintained at development time, essential parts from the goal models are possible to transform into a simple process model skeleton. The STS-Tool is not only, as previously described, able to output a textual description of the security requirements and threats. The STS-models can also be output to a machine-readable security requirements specification (SRS) format for use in the transformation. Although there is no way to completely automate the transformation from STS to BPMN, a transformation tool can provide a structured way to manually support selection of threats that are relevant for the service process model~\cite{aniketosD52}, as shown in Figure~\ref{fig:mtm-transformation}.

Transformation of STS-ml threats into BPMN threats is preferable in cases where we have input from monitors that can give the modelled threats a meaningful role at runtime. An advantage of the tool-supported model transformation is that the threat ID, which enables linking a modelled threat to the threat repository, can be conveniently transferred between the models. In that way, any further transformation and use of the threat can be done with support from the threat repository, and the information always to be found here.

In our case study, we have \textit{Unavailable component} as a threat in the goal model, which can be specialised into e.g.\ a \textit{DDoS attack on service component} threat in the service process. This threat can be fairly easy to monitor, and we hence are potentially able adapt our service in case this threat escalates.

\subsubsection{Threat Response Recommendations}
A separate component, also integrated with the online threat repository, implements logic and knowledge to find and recommend possible mitigations for threats~\cite{aniketosD43}. These countermeasures can potentially be provided in various formats, ranging from textual descriptions to actual code or services.

\begin{figure}[htb]
    \centering
    \includegraphics[width=1\textwidth]{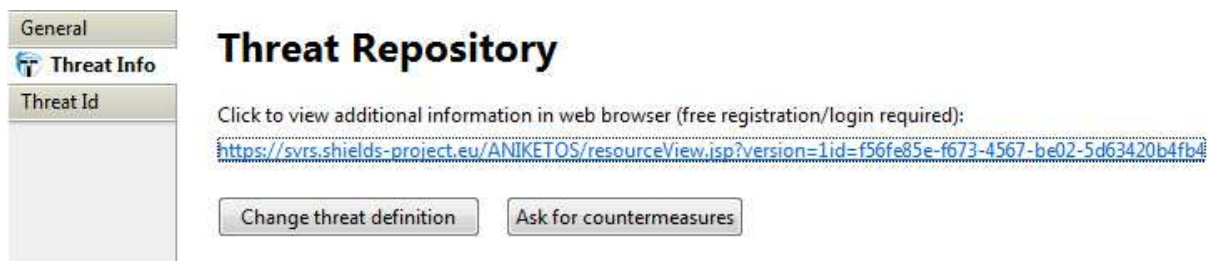} %,bb=0 0 722 150 
  \caption{Threat ID is maintained so the service process modelling tool can ask for countermeasures.}
\label{fig:countermeasures}
\end{figure}

In order to obtain the threat response recommendations, the threat IDs are gathered from the BPMN diagram and the relevant \textit{ErrorBoundaryEvent} elements. Again, the threat ID is input to the threat repository, and returned are the countermeasures. As there may be several countermeasures available for each threat, the various options are ranked before being presented to the service designer.

\subsubsection{Rules Definition}
Not only finding appropriate countermeasures, but actually implementing those that are appropriate, is essential to improve security of the service. Based on threats defined in the diagram and any other events that are monitored and one can be notified about, rules can be defined to address the scenarios we are able to foresee. Rules assume some kind of monitoring in place to provide actual value, but as long as the format can match what the SRE is able to interpret from the individual monitors, the rules are in practice agnostic to monitoring implementations.

\begin{figure}[htb!]
    \centering
    \includegraphics[width=1\textwidth]{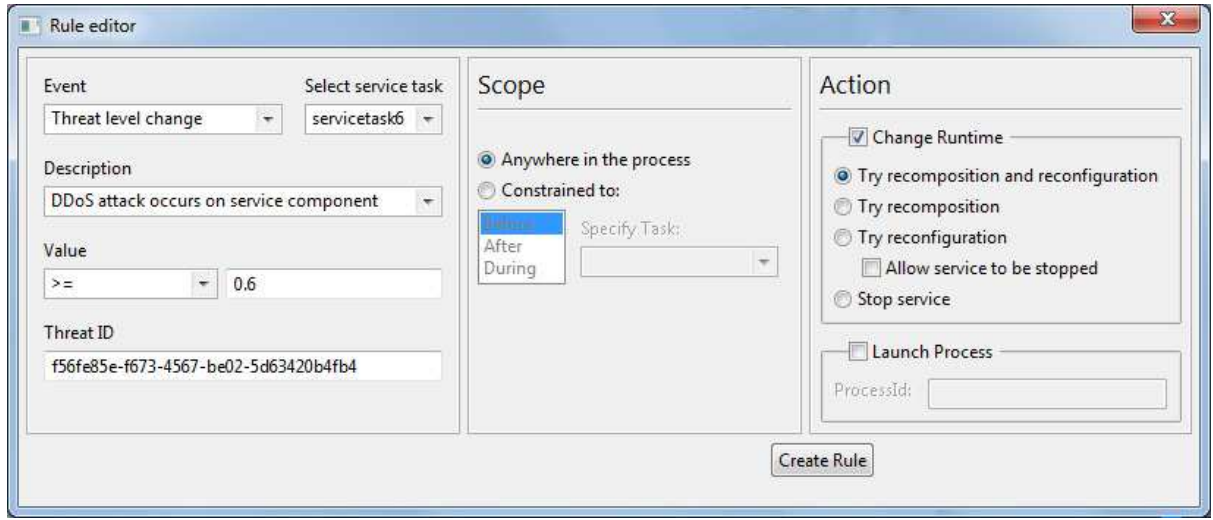} %,bb=0 0 821 350 
  \caption{Threat ID appears in rule editor when threatened task is selected as basis for a new rule.}
\label{fig:rule-editor}
\end{figure}

The service tasks form the basis for defining such rules, as shown in Figure~\ref{fig:rule-editor}. The tasks are both accurate in targeting the rule's scope, yet independent of the actual service implementation we choose. Therefore, we do not need to define individual rule-sets for each potential composition plan. 

\begin{figure}[htb]
    \centering
    \includegraphics[width=1\textwidth]{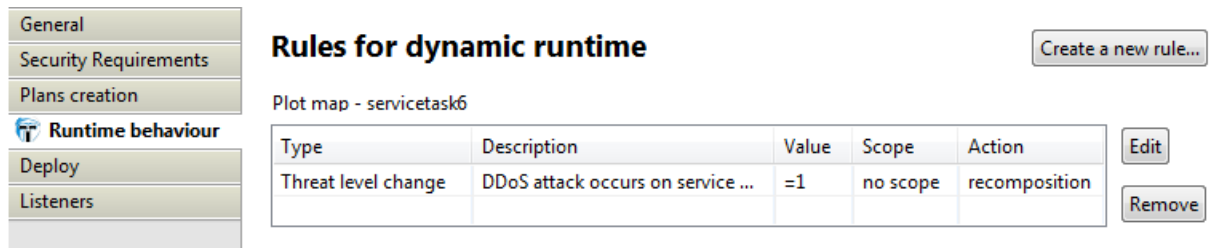} %,bb=0 0 753 150 
  \caption{A rule has been defined for responding to a DDoS-attack on the map service.}
\label{fig:rule}
\end{figure}

In our case study service process we have already modelled a \textit{DDoS-attack occurs on service component} threat, and this can be addressed by choosing \textit{Threat level change} as the event type, and then selecting the service task it applies to. For values in the rule, we choose a decimal number between 0 and 1 (inclusive). This matches our threat monitoring module, which can send alerts with probability value for the escalation of threats~\cite{aniketosD43}. Further, in the scope section of the rule editor, it is possible to define where in the process the rule shall apply. If a particular threat escalates for a component, we might not need to perform reactive measures unless it happens before, after or during the execution of a particular task in our process. Anyhow, we have several options for also defining the action to perform when the rule is matched, such as simply stopping the service execution (and further provision), or trying to recompose or reconfigure the service. We also have an option to launch an additional service process, e.g.\ one that might initiate hardening of other parts of the system, and/or send notification messages to clients and/or service technicians. In the end, we might end up with a list of several rules for several service tasks, or simply one for the one we have defined in our case study, as shown in Figure~\ref{fig:rule}.

\subsection{Runtime Management}

%\subsubsection{Detection}
%Monitors can have all kinds for forms and shapes, and likewise output alerts they produce in various formats.

\subsubsection{Notifications}
In order to operate a self-adapting service infrastructure at runtime, a commonly supported system for machine-to-machine (M2M) messaging is needed. Since there may be an endless number of services and SREs utilising this infrastructure, we cannot provide all messages to everyone. A publish-subscribe pattern is rather suitable, since we already define rules for what we need to respond to. Hence, appropriate subscriptions can be derived automatically from these rules and registered by the SRE. The SRE will then receive notifications only as specified, although the granularity of the rules will determine the relevance of e.g.\ slight variations in threat level probabilities (not all changes will actually trigger an action).

The SRE in our proof-of-concept creates subscriptions based on the rules which are attached to service deployment. It receives notifications according to the subscriptions, and whenever such notifications arrive, all rules are being checked for a match. The messaging system is based on Apache ActiveMQ~\cite{activemq}, which utilises the Java Messaging Standard (JMS) to provide compatibility for many platforms and alternative protocols. ActiveMQ subscriptions are registered with a centralised broker, or network of brokers for horizontal scalability, since it is the broker(s) that deal with receiving and dispatching the notifications to all subscribers. Our broker is in addition deployed on a cloud-based infrastructure, for further increased scalability. In addition to threat level changes, we have implemented support for notifications about changes in trustworthiness, service contract violations, security properties of service components, service runtime context, and service component changes. The notifications are delivered on a best-effort basis, and may of course arrive too late in some cases (depending on how quickly a problem is discovered and duration of the attack).

\begin{figure}[htb!]
    \centering
    \includegraphics[width=1\textwidth]{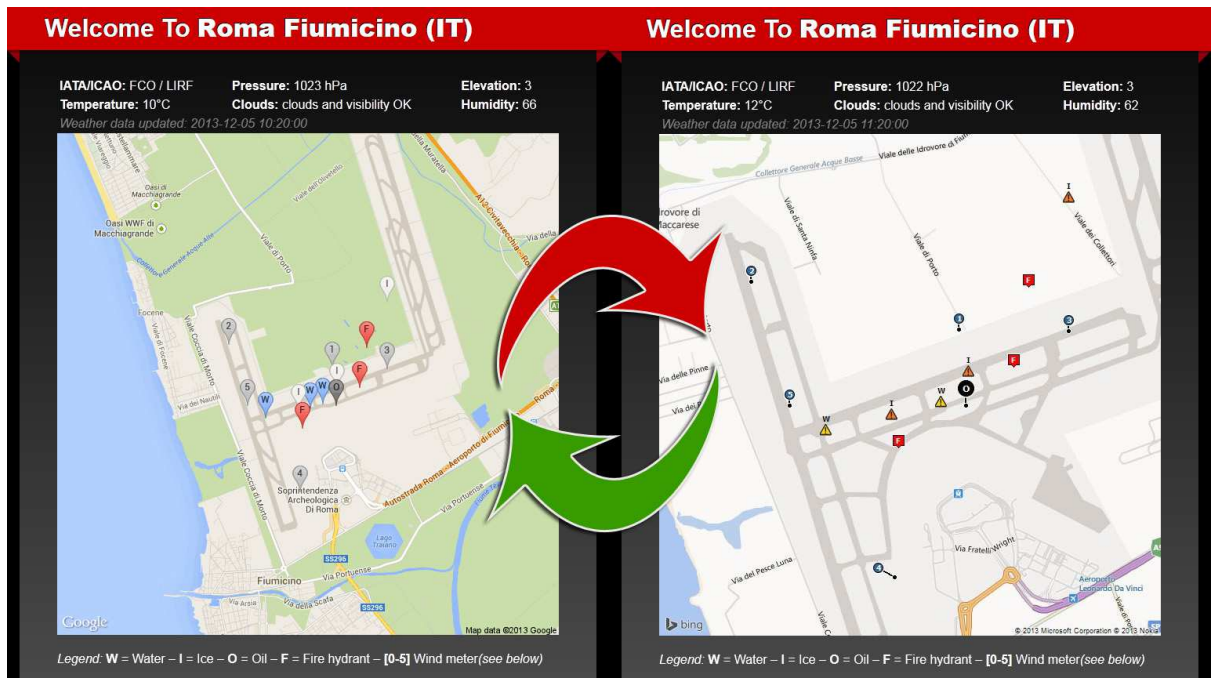} %,bb=0 0 1536 856 
  \caption{The SRE has replaced the Google map service with Bing Maps through a recomposition.}
\label{fig:recomposition}
\end{figure}

\subsubsection{Dynamic Adaptation}
Since a threat monitor has now purportedly detected that the original map service used in our case study is hit by a DDoS-attack, the SRE becomes notified about this through the notification service. In order to trigger events in the self-configuring service process, the SRE must naturally be able to receive such notifications and align these with the previously defined and deployed rules. As it finds a match with the rule we defined earlier on DDoS-attacks, the SRE initiates the specified action according to that rule, which is here to try a recomposition. 

Since we had support for two different map services providing the same functionality, we have prepared an additional composition plan for the airport report service. When the notification concerning a DDoS-attack on the map service is triggered and received by the SRE, the first plan no longer satisfy our security requirements through verification~\cite{zhou2012secure}. Since the rule in Figure~\ref{fig:rule} provides the match, a recomposition is initiated accordingly. The original composition plan with the DDoS-ed map service will be ignored, and the second plan becomes the top-ranked one based on a chosen ranking criteria. The recomposition proceeds with deploying the second composition plan, containing the alternative map service instead of the original one. Nevertheless, the same level functionality is provided, as illustrated with Figure~\ref{fig:recomposition} where the airport reports, before and after recomposition, are lined up next to each other. Before the new composition is deployed, the SRE invokes a runtime verification, and the time this takes depends on the complexity of the composition. This is a well-known scalability issue associated with any kind of runtime adaptation.

% Edit separate file main.tex

\section{Discussion}
\label{discussion}
An obvious shortcoming to our approach is that we do not integrate a proper information security risk assessment anywhere. This is usually where threats are actually identified in a systematic manner, and where they are ranked in the order of overall risk. With this approach, we do assume that such risk assessment activities are already taking place, and that the results from these are actively used to enrich the goal and service process models. However, we also suggest that enrichment can also occur in the opposite direction, so that threat discoveries made during the requirements engineering and service design process can indeed enrich the identification phase of a risk assessment. If tool-support is to be added for this interactivity, it is indeed possible to use the threat repository service for exchanging the threat information between applications.

In terms of threats within the service process model, we have been forced to make a pragmatic compromise. Since BPMN is a process language, it is capable of defining an entire process as it is \textit{intended} to flow, however not without language constructs for handling quite a few exceptions. Threats also provoke exceptions to normal flow, and although explicitly missing from the language, we have earlier concluded that the graphical language itself is even capable of representing what is needed here~\cite{conf/IEEEares/MelandG12}. Still, we found that runtime support for handling externally triggered notifications on the process level is not handled in BPMN software tools such as the Activiti process engine. That is the reason we developed a custom plug-in for the SRE, and consequently a compatible rule editor was developed and integrated into the BPMN tool as well.

Another limitation with the tool-set we present, is that process diagrams are not supported with automatic reflection of changes in the goal diagrams. This may impact an iterative process which goes back and forth between goals and process models. Mitigating this problem through re-transformation is possible, albeit not very appealing, since any process information that is not part of the goal diagram and its security commitments will be lost. Our transformation tool is however capable of comparing any BPMN-diagram with a security requirements specification document, in order to point out nonconformity with the original security requirements. While not yet implemented, this tool could in theory be made capable of also including threats in the analysis, and point out which threats are missing in the BPMN-diagram. The problem here is however if a business-level threat from the goal model, such as ``Integrity error'', has been manually sub-typed by the process designer to either one or several operational-level threats, e.g. ``Checksum mismatch'' or ``Certificate verification failed''. Our threat repository does not express a hierarchy of the threats, and we cannot automatically trace neither the parent from a child nor the children of a parent. It is not always straightforward to create such threat hierarchies either, in particular when a child has several parents.

The main motivation for threat modelling described in this paper is related to preventive and corrective measures for design-time creation and runtime execution of composite services. Another significant motivation that we have not delved into here is the need for engineers and business-people to communicate better around security when systems are designed and implemented, as pointed out by Wolter et al.~\cite{wolter}: \textit{``It is evident that both security and business domain experts need to be able to define their security goals collaboratively on a common abstraction level''}. For this we believe that the use of threats in well-defined graphical modelling languages, such as STS-ml and BPMN can be a positive contribution. We currently have ongoing work on practical evaluation of the presented tools and services within the ATM domain (and others), along with other functionality on security requirements management which is out of scope for this paper. Through this evaluation we hope to gain more knowledge on what will be necessary for a broad uptake of such an approach to managing threats. Along with the already available STS-Tool~\cite{ststool}, the described software modules will eventually be released as part of our extended BPMN tool, and/or as online services.

% Edit separate file discussion.tex

\section{Conclusion}
\label{conclusion}
We have demonstrated how graphical threat modelling can benefit design-time mitigation and runtime correction of unwanted incidents. Though a service may be regarded as secure enough in the early life-cycle phases, risks are not static, and threats, vulnerabilities, probabilities and consequences can change abruptly. Tool-support is therefore essential to be able to analyse which measures to implement at design-time, and create mechanisms to handle the residual ones at runtime through e.g. automatic adaptation.

An essential part of the tool-chain we have presented is the use and re-use of threats through a shared threat repository that many tools are able to interface with. Having a common reference on the name, nature and mitigation options for the threats has been valuable to us both during the graphical modelling at design-time and for runtime alert messaging and response. By basing our work on existing languages and tools, we believe that chances of uptake is significantly larger compared to a comprehensive tool-set built from scratch. However, this also means that some transitions between the tools can be seen as a bit cumbersome. For our future work we seek to gain more knowledge on usability and integration with existing risk assessment methods that should complement the threat modelling.
% Edit separate file conclusion.tex

\section{Acknowledgement}
The authors would like to thank in particular Francesco Malmignati, Balazs Kiss, Mauro Poggianella, Mattia Salnitri, Eider Iturbe and Konstantinos Giannakakis for their technical work related to enabling our proof-of-concept implementation. The research leading to these results has received funding from the European Union Seventh Framework  Programme (FP7/2007-2013) under grant no 257930 (Aniketos).

%\section{Bibliography}
%\nocite{*}
\bibliographystyle{eptcs}
\bibliography{ThreatsLifecycle_GraMSec2014}

\end{document}